\documentclass[aps,showpacs,twocolumn]{revtex4}
\usepackage{amssymb}
\usepackage{graphicx}
\usepackage{dcolumn}
\usepackage{amsmath}

\begin{document}

\title{Electronic States and Anomalous Hall Effect in Strongly Correlated Topological Systems}

\author{Valentin Yu. Irkhin}
\affiliation{M.N. Mikheev Institute of Metal Physics UB RAS, 620108, S. Kovalevskaya str. 18, Ekaterinburg, Russia}
\email{Valentin.Irkhin@imp.uran.ru}

\author{Yuri N. Skryabin}
\affiliation{M.N. Mikheev Institute of Metal Physics UB RAS, 620108, S. Kovalevskaya str. 18, Ekaterinburg, Russia}

\begin{abstract}
We treat elementary excitations, the spin-liquid state, and the anomalous Hall effect (including the quantum one in purely 2D situation) in layered highly correlated systems. The mechanisms of the formation of a topological state associated with bare flat energy bands, correlations, and spin-orbit interactions, including the appearance of correlated Chern bands, are analyzed. A two-band picture of the spectrum in metallic kagome lattices is proposed, which involves a transition from the ferromagnetic state, a flat strongly correlated band, and a band of light Dirac electrons. In this case, the effect of separation of the spin and charge degrees of freedom turns out to be significant. The application of the representations of the  Kotliar-Rukenstein auxiliary bosons and the Ribeiro-Wen dopons  to this problem is discussed.
\end{abstract}

\maketitle

like the quantum Hall effect

in a strong magnetic field

zen enlightenment

happens instantly

leads along the way in steps

between plateaus

and is closely related

with burden of inhomogeneities

and edge states

it never gets around

without impurities and defects

and other pitfalls of practice

which are irrelevant

for quantization accuracy

\section{Introduction }

Recently, a number of layered compounds with competing ferro- and antiferromagnetic phases have been intensively studied, including systems with frustrated (triangular, honeycomb, and kagome) lattices, which exhibit anomalous quantum Hall effect (QHE). For example, this effect is observed [1] in the antiferromagnetic topological insulator MnBi$_2$Te$_4$ with ferromagnetic triangular layers [2]. Of particular interest are systems with a ferromagnetic ground state and flat bands, where Dirac electronic states arise, which can lead to topological Chern insulator phases. Such states were observed in a number of layered compounds of transition metals with a kagome lattice Fe$ _3 $Sn$ _2 $, Fe$_3 $GeTe$ _2 $, Co$_3 $Sn$ _2 $S$ _2 $,  FeSn, etc. (see discussion in [3--8]). Recently, a ferromagnetic Chern insulator state with a large anomalous Hall effect has been observed in the moire structure of three-layer graphene [9]. The anomalous QHE is also observed in bilayer graphene [10]. 

In the systems under discussion, one can expect the formation of exotic topological quantum states. Unusual excitations arising in two-dimensional strongly correlated systems may obey nonstandard statistics, including fractional statistics. The history of the study of these excitations started from the fractional QHE [11], which means the topological state of matter. In this case, low-energy physics is described by the Chern--Simons gauge theory. 

Topological Hall phases can also appear in lattice systems in the absence of an external magnetic field (the anomalous Hall effect). The realization of such phases on a lattice requires a number of conditions to be met. First, this is the presence of a flat (almost dispersionless) bare energy band with nontrivial topology (nonzero Chern number, which determines the number of edge excitations), which makes it possible to implement physics similar to the Landau level picture. The second important condition is a strong electron--electron interaction, which violates the Fermi liquid picture. Strong correlations are especially important for the fractional Hall effect, when the Landau levels are highly degenerate. 

Since the Hall conductivity is odd with respect to time reversal, topologically nontrivial states can arise when the corresponding $T$ symmetry is broken. Thus, one of the possible mechanisms is the spontaneous breaking of this symmetry due to coupling to vortices [3, 12]. 

In a number of works, attempts have been made to take into account strong interactions in a highly frustrated system in order to obtain a phase with topological order in simple mean-field-type approximations. For example, the anomalous QHE can be dynamically generated in a generalized Hubbard model on a honeycomb lattice and in other lattice systems with a quadratic band crossing point, including a kagome lattice. However, detailed numerical studies do not confirm the formation of exotic topological phases predicted by mean-field theories. The main difficulty here is that, instead of triggering the spontaneous $T$ symmetry breaking, strong interactions also tend to stabilize competing long-range orders that break translational symmetry (see discussion in [12]). Thus, when describing lattice systems with the QHE, it would be appropriate to start directly with a strongly correlated state. 

The formation of a fractional quantum Hall state was also considered in Hubbard-type models involving chirality [13, 14]. 

The presence of spin--orbit interaction admits another topological class of insulating band structures, in which the $T$ symmetry is not broken initially [15]. Such a two-dimensional topological insulator is called a quantum spin Hall insulator. This system is described by the double Haldane model [16] with opposite signs of the Hall conductivity for up and down spins. In an applied electric field, the up and down spins give Hall currents that flow in opposite directions. Thus, the total Hall conductivity is zero, but there is a spin current and a quantized spin Hall conductivity. When the complex amplitudes of hopping between next-nearest neighbors are taken into account, a gap opens at the Dirac points, and the @T symmetry is broken, which leads to two bands with Chern numbers of $C= \pm 1$ and to the integer QHE in the case of half filling. For flat bands with certain fillings, a topological state of a quantum incompressible liquid---the fractional QHE---arises [17]. 

There is also a third possibility, which was considered in [18, 19] in relation to bilayer graphene and seems to be the most interesting case. For the Chern spin band with valley polarization, the class of exotic phases of Chern insulators is described in terms of the separation of spins and charges: the charges are in the phase of an ordinary Chern insulator with quantized Hall conductivity, and the spins form a disordered spin-liquid phase, a quantum Hall spin liquid, similar to the usual $Z_2$ spin liquid with fractional degrees of freedom. The condensation of spinon quasiparticles in such a Hall spin liquid can lead to a quantum Hall antiferromagnet. 

Two-dimensional and three-dimensional topological systems exhibit a number of unique properties. In particular, Dirac and Weyl topological semimetals represent a new class of topological materials. Relativistic fermions---low-energy excitations around Dirac and Weyl points---lead in such materials to exotic transport characteristics, including high magnetoresistance and intrinsic anomalous Hall effect [20]. In this paper, we consider the relationship between electronic states and the anomalous Hall effect in strongly correlated topological systems. We will try to demonstrate that the physical situation in such systems has a number of interesting differences from more conventional topological systems (see, for example, [21, 22]). 

In Section 2, we consider the Chern states for the Landau level and for lattice models in systems with correlations. In Section 3, we turn to metallic states with special emphasis on kagome lattices and discuss the application of an effective two-band model, which includes a narrow Chern band and a wide carrier band, to these systems. In Section 4, we analyze in more detail experimental data, including data on the Dirac and Weyl metals and semimetals. 

\section{ Chern states } 

In the QHE situation, electrons in a magnetic field move circularly along cyclotron orbits, including around other electrons. In this case, the Landau level number is determined by the number of wavelengths on a given circle. The description of strongly correlated systems includes the formation of flat bands, which are analogous to the Landau levels. In this case, the total Berry flux over the Brillouin zone is (the sum of the integrals of the Berry curvature over all occupied bands [15]) is a Chern topological invariant $C$. This is how the state of the Chern insulator is formed, the distinctive feature of which is the presence of gapless edge states. 

The fundamental property of gapped topological systems is the existence of gapless conduction states at interfaces where the topological invariant changes (in the simplest case, at the boundary with empty space). These states can be understood as a consequence of the cyclotron orbit bounce off the edge. It is important that the electronic states responsible for this motion are chiral: electrons propagate in only one direction along the edge. Such states are insensitive to disorder since there are no states available for backscattering. This fact underlies the perfectly quantized electronic transport in the QHE [15]. Topological surface states are also stable with respect to Anderson localization. 

In the presence of interaction, a system can be considered within the Hubbard model with a bandwidth $W$ and Coulomb repulsion $U$. In the case of the Chern band, due to the obstruction of the Wannier states, the charge cannot be localized, so that the narrow band regime is not described by a purely spin model [19]. Therefore, in topological strongly correlated systems with Chern bands, localized states are absent, and there is no ordering of localized moments---magnetism can only be itinerant. 

For small $W/U$ ratios, the charge gap should be determined by the value of $U$, and the spin disorder does not necessarily close the charge gap. In a topologically trivial band with nonzero Berry curvature, ferromagnetism in the limit of $W = 0$ can be suppressed by antiferromagnetic exchange [18]. A similar destruction of quantum Hall ferromagnetism by the kinetic term is also possible at integer filling of the Chern band. For intermediate values of $W/U$, the charge subsystem remains a Chern insulator with quantized Hall conductivity, but the spins are not in a ferromagnetic state. Thus, quantum Hall antiferromagnetic and quantum Hall spin-liquid phases arise in which antiferromagnetic order or a quantum spin liquid coexist with quantized Hall conductivity [19]. Thus, a transition from a ferromagnetic state with the QHE to a Fermi liquid through unusual antiferromagnetic and spin-liquid phases is possible. 

The effective Lagrangian describing the QHE for electrons in a magnetic field and including the Chern--Simons term has the following form [23, 24]: 
\begin{equation}
	{\cal L}_{CS} = -
	\frac m {4\pi} \epsilon^{\mu\nu\lambda}a_{\mu}\partial_\nu a_{\lambda}
	- \frac e {2\pi} \epsilon^{\mu\nu\lambda} A_{\mu}\partial_\nu a_{\lambda},
	\label{eq:C2}
\end{equation} 
where ${ a}_\mu$ is the internal gauge field, ${ A}_\mu$ is the vector potential of the external electromagnetic field, and $\epsilon $ is a second-rank antisymmetric tensor. The filling is $\nu=1/m$, where $m$ is the charge of the gauge field, that is, the number of wavelengths when one electron goes around another. 

Equation (1) describes only the linear response of the ground state to external electromagnetic fields. To have a more complete description of a topological liquid, such as a fractional quantum Hall liquid, one should introduce excitations into the effective theory. Although the electron is a fermion in the Laughlin ground state, the excited states of the system are bosonic. Thus, $m$ is an even integer for bosonic states and an odd number for fermionic states. A fractional quantum Hall liquid contains two types of quasiparticles: a quasihole (or vortex) in the original electron condensate and a quasihole (or vortex) in the new boson condensate. Introducing the gauge field ${\tilde a}_\mu$ describing a bosonic current, we can express the total Lagrangian in a compact matrix form similar to (1): 
\begin{equation}
	{\cal L} = -
	\frac 1 {4\pi} \epsilon^{\mu\nu\lambda}K_{IJ}a_{I\mu}\partial_\nu a_{J\lambda}
	- \frac e {2\pi} \epsilon^{\mu\nu\lambda}q_{I}A_{\mu}\partial_\nu a_{I\lambda}.
	\label{eq:C3}
\end{equation} 
Here  $(a_{1\mu},a_{2\mu})=(a_{\mu},{\tilde a}_\mu)$, the matrix $K$  has the form
\begin{equation}
	K=\left(
	\begin{array}{cc}
		p_1 &-1\\
		-1&p_2
	\end{array}
	\right)
	\label{eq:C_2_K}
\end{equation}
where $p_1$ is the starting number  $m$ describing Fermi electronic states, $p_2$ is an even number describing the bosonic field, $\bf q$ is the charge vector, and ${\bf q}^T = (q_1,q_2) = (1,0)$. For the occupation numbers, we have $\nu = {\bf q}^T K^{-1}{\bf q}$. 

The topological theory of abelian phases of a quantum Hall liquid is generally described by the Lagrangian [23, 25] 
\begin{equation} 
	\label{EFT}
	\mathcal{L}_{\text{bulk}}=\frac{1}{4\pi} \epsilon^{\mu\nu\rho} a^I_{\mu} K_{IJ}\partial_\nu a^J_{\rho}
	- a^I_{\mu} j_{I}^{\mu}
	-\frac{1}{2\pi} t_{AI} \epsilon^{\mu\nu\rho}\mathcal{A}^A_{\mu}\partial_\nu a^I_\rho.
\end{equation} 
Here $a^I$ ($I=1,2,\dots, N$) is a set of gauge fields, $K_{IJ}$ is an $N\times N$ symmetric integer matrix that determines the mutual statistics of excitations, $j_I$ are quasiparticle currents, and $t_{AI}$ is a charge vector that determines the occupation numbers. The degeneracy of the ground state on the torus (a characteristic of the topological order) is determined by the determinant of the matrix $K$; this determinant also determines the number of independent types of anyons---fractional charge particle. The last term in (4) describes the coupling to external sources $\mathcal{A}^A$  ($A=1,2,\dots, M$) with the global U$(1)_A $ symmetry.  

Incompressible fractional quantum Hall liquids have a finite energy gap for all their bulk excitations. However, such liquids in finite-size systems always contain one-dimensional gapless edge excitations with a complex structure that reflects a rich bulk topological order (the bulk-- boundary correspondence). Thus, it becomes possible to study bulk topological orders by studying the structure of edge excitations. The edge states of non-abelian fractional quantum Hall liquids form even more exotic one-dimensional correlated systems [24]. 

Effective field theory (4) is also well suited for understanding the physics of edge states. In the absence of external sources $\mathcal {A}^ A $, the chiral Luttinger theory arises [25, 26]: 
\begin{equation}
	\label{edge}
	\mathcal{L}_{\text{edge}}=\frac{1}{4\pi}\Big[K_{IJ}\partial_t \phi^I \partial_x \phi^J-V_{IJ}\partial_x \phi^I \partial_x \phi^J \Big].
\end{equation} 
Here $\phi^I$ are $ N $ chiral bosons, $V_{IJ}$ is a nonuniversal positive definite real matrix depending on the microscopic properties of the edge. 

Using the language of axiomatic topological field theory, we can generalize this consideration to non-abelian phases [25, 27]. 

In the case of a strongly correlated electron band, a parton construction---representation of multielectron Hubbard operators $\tilde{c}_{i \sigma }=X_i(0,\sigma )$ in terms of slave (auxiliary) bosons and fermions---is used to describe various classes of quantum Hall spin liquids. This construction may have alternative forms: charged fermionic holons and neutral Bose (Schwinger) spinons (slave-fermion representation) or neutral fermionic (Abrikosov) spinons and bosonic holons (slave-boson representation). In addition, a U(1) gauge field arises due to the constraint on filling at a site. Parton constructions can be introduced in the case of arbitrary symmetry groups corresponding to different types of anyons [28, 29]. They allow one to match the physics of Landau levels and Chern bands, continuous fractional quantum Hall phases, and spin liquids for lattice models, as well as to describe phase transitions at half filling as transitions with a change in the Chern number between insulator phases [30]. 

The mean-field Hamiltonian for the state of a chiral spin liquid is equivalent to the problem of electron transfer in a magnetic field, so that the coupling between slave fermions and the gauge field is identical to the coupling between electrons and the electromagnetic field [24]. Thus, it can be expected that a phenomenon similar to the Hall effect can occur in a system of slave fermions. In the presence of fluxes, quantization in the gauge field gives rise to Landau-type levels [24, 31]. In this case, the zero level has a degeneracy equal to the number of the flux quanta. The addition of a flux quantum generates a zero fermionic mode for each type of Dirac fermions. After switching on the potential of the crystal lattice, the Landau levels turn into narrow correlated bands. In this sense, the Hubbard bands (which can be described in the simplest Hubbard-I approximation, including for degenerate bands [32, 33], as broadened atomic levels) are spinon bands. It can be hypothesized that the Hubbard broadening of levels (for example, due to scattering and resonant broadening [34], as well as due to the interaction of carriers with local moments) plays a role similar to that of disorder, which is essential for the QHE. 
The slave-fermion representation 
\begin{equation}
	\tilde{c}_{i \sigma } = b_{i\sigma} f_i
\end{equation} 
allows one to describe a Z$_2$ spin liquid (singlet pairing of Schwinger bosons) with a spin gap and topological order and, in the Bose condensate regime, also the antiferromagnetic phase (cf. [35]). In this case, the charge response of the Chern insulator is preserved, which means the state of the quantum Hall spin liquid. The construction of this state is generalized to the case of fractional filling and the fractional QHE, so that the quantum Hall Z$_2$ spin liquid gives the realization of eight different abelian topological orders with four anyons. From the viewpoint of topological order, they are equivalent to eight abelian topological superconductors of the 16-fold way [27].

The fermionic spinon representation 
\begin{equation} \tilde{c}_{i \sigma } = b_i f_{i\sigma}
\end{equation} 
describes a U(1) spin liquid with a spinon Fermi surface, which is the parent state for the Z$_2$ spin liquid; this liquid arises when the gauge symmetry is lowered. For even values of the Chern invariant $C$, the so-called composite Fermi liquid state can be constructed, which is analogous to a spin liquid with a purely spinon Fermi surface but is paramagnetic and metallic [19]. This state allows one to partially preserve the electronic degrees of freedom (nonzero residue) and describe the Fermi band. It is important to emphasize that the formation of this state is due to the separation of the spin and charge degrees of freedom in a strongly correlated system. 

The simplest version of the Chern band with an even Chern invariant $C$ is given by the bosonic integer QHE (BIQHE) phase [19]. The effective action for this phase is 
\begin{equation}
	{\cal L}={\cal L}_{bIQHE}[b,A+a]+{\cal L}_{FS}[f,-a].
\end{equation} 
Here $A$ is an external U(1) gauge field; symbolically, in terms of a differential form, we have 
\begin{equation}
	\label{biqhe}
	{\cal L}_{bIQHE}= \frac{C}{4\pi}(A+a)d(A+a),
\end{equation}
\begin{equation}
	\label{fs_a}
	{\cal L}_{FS}=f^\dagger_\sigma(\partial_\tau -\mu+ia_0)  f_\sigma-\frac{\hbar^2}{2m^*}f^\dagger_\sigma(-i\partial_i+a_i)^2f_\sigma ,
\end{equation} 
where $\mu$ is the chemical potential, $m^*$ is the effective mass, and $a_0$ and $a_i$ are the time and coordinate components of the gauge field [23]. Fermionic spinons $ f_\sigma $ partially fill the band without Chern number and form a Fermi surface (FS). 

In the case of odd $C$ = 1, 3, 5, 7... exotic states with the Hall conductivity $\sigma_{xy} = C$ are formed: eight types of spin liquids with a half-integer chiral central charge $c = C-1/2$, which are analogous to eight non-abelian Kitaev superconductors [27]. 

\section{A two-band model for the kagome lattice}

Now, we turn to realistic conducting systems with topological motifs. A two-dimensional layer of the kagome lattice gives a flat band and a pair of Dirac bands, which are protected by symmetry, as in graphene. When the spin--orbit interaction is taken into account, a small forbidden band opens in two-dimensional layered metal compounds with a kagome lattice, and a band with linear dispersion is distorted, so that Dirac fermions acquire a small mass. Unlike linear bands with light quasiparticles, flat bands have no dispersion over a wide momentum interval and behave like Landau levels, which can lead to unusual quantum states, including fractional Hall states. Combined with the spin--orbit coupling and the total magnetization, a phase of a two-dimensional Chern insulator with the QHE is realized at fillings of 1/3 and 2/3. When such layers are superimposed on each other along the third dimension, the interlayer interaction transforms the system into a three-dimensional phase of the Weyl semimetal with broken time reversal symmetry. At the same time, the flat band carries a finite Chern number and mimics Landau levels without an external magnetic field, which allows the implementation of the fractional QHE at a partial filling of flat bands [36]. 

Since the energy gap caused by the spin--orbit interaction of the out-of-plane orbitals is much smaller than that of the in-plane orbitals, one obtains an orbital-selective character of Dirac fermions [7]. Thus, in metals with a kagome lattice, a band of light carriers (Dirac fermions) coexists with a band of heavy electrons (flat band). This leads to an unusual physics if any of these bands comes close to the Fermi level. A detailed calculation of the spectrum was carried out in [7] using the example of a CoSn compound (with Co$_3$Sn layers) with metallic conductivity. 

Here a certain analogy arises with the situation of nodal--antinodal dichotomy in superconducting cuprates, where the spectrum is gapless (electron degrees of freedom are preserved, and excitations are described as Dirac fermions) near the nodal points $(\pm\pi/2,\pm\pi/2)$ of the Brillouin zone and is gapped near the antinodal point $(0,\pi)$ [37, 38]. However, in contrast to cuprates, a significant role is played here by the spin--orbit interaction. 

Depending on the electron--electron and spin--orbit interaction parameters, as well as on the next-nearest neighbor transfer, different arrangement patterns of wide and flat bands arise [39], the bands becoming mixed with decreasing $U$. Thus, we can speak of a Mott-type metal--insulator transition, in which one can expect a change in statistics and a separation of spin and charge. 

An important feature of kagome lattices is the metallic ferromagnetic state, which can transform into a spin liquid by a Kitaev-type mechanism [40]. According to ab initio calculation [41], the doping in the Co$_3$In$_x$Sn$_{2-x}$S$_2$ system preserves half-metallic ferromagnetism with a linearly decreasing magnetic moment. This system also contains Fermi arcs and nodal rings, which play an important role in the anomalous Hall effect [41, 42]. Single-crystal experimental data on the Co$_3$In$_x$Sn$_{2-x}$S$_2$ and Co$_{3-y}$Fe$_y$Sn$_{2}$S$_2$ kagome systems [43] show that these systems have an almost two-dimensional itinerant magnetism and a chiral spin state in the neighborhood of the quantum transition from ferromagnetic to nonmagnetic phase; in addition, a strongly correlated state with a high electronic heat capacity is formed. 

Bands with orbital-selective motifs were also found in the YCr$_6$Ge$_6$ paramagnetic kagome lattice [44] and in the layered FeSn compound with Fe moments ferromagnetically aligned within kagome lattice planes but antiferromagnetically coupled along the $c$ axis [36]. 

A microscopic description can be obtained in terms of auxiliary particles. The most convenient and universal is the rotation-invariant version of the Kotliar-Ruckenstein representation [45]: 
\begin{equation}
	\tilde{c}_{i\sigma }^{\dag }=\sum_{\sigma ^{\prime }}f_{i\sigma' }^{\dag }z_{i\sigma'
		\sigma}^{\dag },~\hat{z}_{i}=e_{i}^{\dag}\hat{L}_{i}M_{i}\hat{R}_{i}%
	\hat{p}_{i}
	\label{zz}
\end{equation}
where scalar and vector bosons, $p_{i0}$ and $\mathbf{p}_{i}$, describing spin excitations are introduced as $\hat{p}_{i}=\frac{1}{2}(p_{i0}\sigma _{0}+\mathbf{p}_{i}\mbox{\boldmath$\sigma $})$,
\begin{eqnarray}
	\hat{L}_{i} &=&[(1-d_{i}^{\dag }d_{i})\sigma _{0}-2\widehat{p}_{i}^{\dag }%
	\widehat{p}_{i}]^{-1/2}  \label{1z1} \\
	\hat{R}_{i} &=&[(1-e_{i}^{\dag }e_{i})\sigma _{0}-2\widehat{\tilde{p}}%
	_{i}^{\dag }\widehat{\tilde{p}}_{i}]^{-1/2}  \label{1z2} \\
	M_{i} &=&(1+e_{i}^{\dag }e_{i}^{{}}+\sum_{\mu =0}^{3}p_{i\mu }^{\dag
	}p_{i\mu }^{{}}+d_{i}^{\dag }d_{i}^{{}})^{1/2}.  \label{1z}
\end{eqnarray}
and $\widehat{\tilde{p}}_{i}$ is the time-reversed operator $\hat{p}_{i}$. In the case of magnetically ordered phases and a low concentration of current carriers (holes), the factors $L$ and $R$ cancel the holon operators $e_i$, and we can approximately write [46] 
\begin{equation}
	\tilde{c}_{i\sigma }=\sqrt{2}\sum_{\sigma ^{\prime }}\hat{p}_{i\sigma
		^{\prime }\sigma }f_{i\sigma ^{\prime }}=\frac{1}{\sqrt{2}}\sum_{\sigma
		^{\prime }}f_{i\sigma ^{\prime }}[\delta _{\sigma \sigma ^{\prime }}p_{i0}+(%
	\mathbf{p}_{i}\mbox{\boldmath$\sigma $}_{\sigma ^{\prime }\sigma })].
	\label{eq:I.88}
\end{equation}
This representation allows one to construct an interpolation between the standard slave-boson representation, where the current carriers are Bose holons, and the spin-wave representation in the magnetically ordered phase. Thus, different scenarios arise for the transitions from the Hall magnetic phase to spin-liquid and then to Fermi-liquid phases upon changing the interaction or doping parameters (cf. [19]). Thus, assigning the Chern number $C$ to these bound states (charge degrees of freedom), one can use the results of [19] to describe the effect of the Chern number on the topological order near certain points of the Brillouin zone. In this case, it is natural to assume that the current carriers inherit the $C$ numbers for the original Chern band (for example, in the mean-field approximation for auxiliary bosons). 

Representation (15) can be formally associated with the dopon representation, where for the Hubbard operator we have [47--49] 
\begin{equation}
	X{_i(0, -\sigma) }=-\frac \sigma{2\sqrt{2}}\sum_{\sigma ^{\prime }}d_{i\sigma
		^{\prime }}^{\dagger }(1-n_{i,-\sigma ^{\prime }})
	[\delta _{\sigma \sigma ^{\prime }}-2(\mathbf{S}_i\mbox{\boldmath$\sigma $}_{\sigma ^{\prime }\sigma} )].
	%+\frac {{(\mathbf{S}_i\mbox{\boldmath$\sigma $}_{\sigma ^{\prime }\sigma })}} {2S+1}\right].
	\label{eq:I.8}
\end{equation}
Here $\sigma=\pm$, $n_{i\sigma}=d^{\dagger}_{i\sigma}d_{i\sigma}$, the Fermi operators of dopons $ d_ {i \sigma ^ {\prime}} ^ {\dagger} $ describe current carriers, and the spin operators $ \mathbf {S} _i $ describe localized degrees of freedom; they can be represented in terms of Fermi or Bose (Schwinger) spinons [35, 48]. In the representation of pseudofermions, we have 
\begin{equation}
	\mathbf{S}_i=\frac 12\sum_{\sigma \sigma ^{\prime }}f_{i\sigma
	}^{\dagger }\mbox{\boldmath$\sigma $}_{\sigma \sigma ^{\prime
	}}f_{i\sigma ^{\prime }}.
\end{equation}

Taking into account the hybridization between dopons and Fermi spinons $f_{i\sigma}$ provides a description within the effective two-band model [48]. Substituting (16) into the $ t-J $ Hamiltonian, we obtain a Hamiltonian with chiral trends (that contains vector products) [48, 49]: 
\begin{multline}
	H=\frac{1}{(2S+1)^{2}}\sum_{ij\sigma \sigma ^{\prime }}t_{ij}\{(S^{2}+%
	\mathbf{S}_{i}\mathbf{S}_{j})\delta _{\sigma \sigma ^{\prime }}-S(\mathbf{S}%
	_{i}+\mathbf{S}_{j})\mbox{\boldmath$\sigma $}_{\sigma \sigma ^{\prime }} \\
	+i\mbox{\boldmath$\sigma $}_{\sigma \sigma ^{\prime }}[\mathbf{S}_{i}\times
	\mathbf{S}_{j}]\}c_{i\sigma }^{\dagger }(1-n_{i,-\sigma })(1-n_{j,-\sigma
		^{\prime }})c_{j\sigma ^{\prime }}+H_{d},  \label{eq:I.10}
\end{multline}
where $ H_ {d} $ is the Heisenberg Hamiltonian and $S = 1/2$. Such a representation of the Hamiltonian in the narrow-band $ s-d $ model with arbitrary spin $S$ was obtained in [47]. For noncoplanar magnetic structures, the vector product term can lead to the anomalous Hall effect even without spin--orbit interaction, due to the spin chirality and Berry curvature. 

The Heisenberg part of the Hamiltonian can be considered in the slave-boson representation. To this end, we can rewrite expression (16) in terms of the operators 
\begin{equation}
	b_{1i}=f_{i\uparrow}d_{i\downarrow}-f_{i	\downarrow}d_{i\uparrow}, \,
	b_{2i}=f_{i\uparrow}^\dag d_{i\uparrow}+f_{i	\downarrow}^\dag d_{i\downarrow},
\end{equation}
which can be approximately considered as Bose operators [48]; this brings us back to the representation of bosonic holons, including in the SU(2) version [37, 50]. 

The mean-field theory [48] includes fermionic spinons and correlated electrons, dopons, the bosonic holons being bound states of spinons and electrons. Note that, in the approach of [47, 49], electron operators appear instead of new operators of dopons, which somewhat changes the physical interpretation (in particular, this may be important for describing the QHE). 

Thus, the original model takes the form of an effective two-band model, similar to the Kondo lattice problem: a flat band is described in the representation of Abrikosov fermions, and the conduction band, in terms of dopons. Here, a (partial) orbital-selective Mott transition occurs in one band, which represents a quantized change in the Fermi surface, i.e., a transition from a large to a small Fermi surface [51], associated with the formation of Hubbard subbands [31, 38, 52]. This transition can be interpreted within the fractionalized Fermi-liquid approach, which describes a small Fermi surface in the state of a Z$_2$ spin liquid [53]. 

In the mean-field approximation, the Lagrangian for the spin-liquid phase can be written as (cf. [48]) 
\begin{align}
	{\cal L}&=
	\sum_{\bf{k \sigma}} \left[
	(\partial_0+\alpha_{\bf{k}}) f_{\mathbf{k}\sigma }^{\dagger }f_{\mathbf{k}\sigma } +
	\beta_{\bf{k}} (f_{\mathbf{k}\sigma }^{\dagger }d_{\mathbf{k}\sigma }+ {\rm h.c.})\right. \notag\\
	&\left. + (\partial_0+\gamma_{\bf{k}}) d_{\mathbf{k}\sigma }^{\dagger }d_{\mathbf{k}\sigma} + {\rm const} \right],
\end{align}
where $\gamma_{\bf{k}}$ is determined by the (generally speaking, renormalized [48]) bare spectrum, $\alpha_{\bf{k}} $ is, besides that, proportional to the concentration of current carriers (dopons), and $\beta_{\bf{k}} $ contains the bosonic renormalizations $\langle f_{i\sigma}^\dagger d_{j\sigma}\rangle$. Thus, after diagonalizing the spectrum, we obtain a narrow $f$-type band and a wide band originating from dopons. Further, we can take into account fluctuations by introducing the time and coordinate components of the gauge field, so that $\partial_0 \rightarrow \partial_0 + ia_0$, and, accordingly, for the coordinate component (cf. [28]). 

A similar approach for the moire lattice of graphene with regard to the valleys makes it possible to describe the formation of a spin-liquid state with a fractional charge of anyons [28]. In general, there are two sources of topological order in our problem. The first is the quantum Hall effect; it can be described within the parton representation for the electron operator [29]. The second is the formation of a spin liquid, where the separation of charge and spin in the strong correlation regime is significant. In principle, both of these factors are described within the parton mean-field theory [30]. It should be expected that the effective Lagrangian of a strongly correlated system would contain terms describing the interference of these effects, which makes it possible to assign the system to a certain topological type. The latter can be defined according to Kitaev's classification of abelian and non-abelian topological superconductors [27].

\section{ Discussion}

3d-metal compounds with kagome lattice are characterized by both topological electronic bands and a variety of magnetic structures. The combination of these two factors can lead to large values of the anomalous Hall conductivity through various mechanisms. In particular, in magnetic Weyl semimetals with broken $T$ symmetry with respect to time reversal, a strong intrinsic anomalous Hall effect arises due to the large Berry curvature. 

In a kagome lattice, the $T$ symmetry can be broken by magnetic fluxes due to ferromagnetic ordering and intrinsic spin--orbit interaction, which gives rise to several nontrivial Chern bands separated in energy and in the intrinsic anomalous QHE. This mechanism was considered in [54] as applied to the compound Cs$_2$LiMn$_3$F$_{12}$. The strong-coupling model used in that paper is similar to the Haldane model and includes two upper bands with dispersion and a lower flat band, which are separated in energy and carry the Chern numbers --1, 0, and +1. The first two bands are linearly touching at the points $\bar{K}$ and $\bar{K}'$, forming two Dirac cones, and the middle and lower flat bands are tangent quadratically at the point $\Gamma$. 

The corresponding tight-binding Bloch Hamiltonian with allowance for the spin--orbit interaction for a kagome lattice, which allows one to construct states with different Chern numbers, was obtained in [55]. This makes it possible to construct a mean-field approximation in the dopon representation similar to [48]. 

Magnetic Weyl semimetals and metals can potentially realize the anomalous QHE in the two-dimensional limit. It can be expected that the structure of the kagome lattice in combination with the interplanar ferromagnetic order in the layered magnetic system Co$_3 $Sn$ _2 $S$ _2 $ would allow the observation of a quantum anomalous Hall state in the two-dimensional limit [3, 56, 57]. Since magnetic Weyl semimetals are topological systems, the anomalous QHE state can be obtained in them due to the confinement of the Weyl semimetal along a certain direction. This idea was developed in [56] using the example of Co$_3 $Sn$ _2 $S$ _2 $. In the two-dimensional limit, two anomalous QHE states were obtained depending on the layer stoichiometry. One of them is a semimetal with a Chern number of 6, and the other is an insulator with a Chern number of 3. 

Note that the Co$_3 $Sn$ _2 $S$ _2 $ compound is a representative of half-metallic ferromagnets, which are characterized by strong correlation effects and in which an important role is played by incoherent (nonquasiparticle) states [58], which may be related to topology [46]. The possibility of the anomalous QHE is also discussed for the three-dimensional half-metallic ferromagnet HgCr$_2$Se$_4$, in which, according to the band calculation [59], the electronic spectrum includes Weyl fermions. 

The approach of [19] allows us to propose a new class of topological phases---quantum Hall spin liquids---which are a combination of a quantum Hall state and a spin liquid, as well as the parent state for quantum Hall ferro- and antiferromagnets; in the case of a kagome lattice, a direct transition is possible from a ferromagnet to a spin liquid. It should be noted that the nonbipartite kagome lattice is favorable for the formation of a spin liquid, while the identification of the lattice type of bilayer graphene from the (bipartite honeycomb or non-bipartite triangular) charge and spin density distributions is not unambiguous [60]. 

Thus, we see that exotic phenomena in narrow topological bands are due to correlation effects, including the separation of spin and charge degrees of freedom. At the same time, the spin--orbit interaction and ferromagnetism play an important role in the formation of flat bands and for the anomalous QHE. 

The work was carried out within the state assignment of the Ministry of Science and Higher Education of the Russian Federation (project ``Flux'' no. AAAA-A18-118020190112-8). The study of spin--orbit interaction effects in kagome lattices was supported by the Russian Science Foundation under grant no. 20-62-46047.


\begin{thebibliography}{99}
	
	\bibitem{MnBi2Te41}
	%Quantum anomalous Hall effect in intrinsic magnetic topological insulator MnBi2Te4
	Yu. Deng, Y. Yu, M. Zh. Shi, Zh. Guo, Z. Xu, J. Wang, X. H. Chen, Yu. Zhang,
	Science  \textbf{367},  895 (2020).
	
	
	\bibitem{MnBi2Te4}
	%The antiferromagnetic (AF) compound MnBi2Te4 is suggested to be the first realization of an antiferromagnetic (AF) topological insulator. that possess ferromagnetic (FM) triangular layers with AF interlayer coupling.
	%Competing magnetic interactions in the antiferromagnetic topological insulator MnBi2Te4
	B. Li, J.-Q. Yan, D.M. Pajerowski, E. Gordon, A.-M. Nedic, Y. Sizyuk, L. Ke, P.P. Orth, D. Vaknin, and R.J. McQueeney,
	Phys. Rev. Lett. \textbf{124}, 167204 (2020).
	
	
	
	\bibitem{Liu}
	E. Liu, Y. Sun, N. Kumar, L. Muechler, A. Sun, L. Jiao, Sh.-Y. Yang, D. Liu, A. Liang, Q. Xu, J. Kroder, V. Seuss, H. Borrmann, Ch. Shekhar, Zh. Wang, Ch. Xi, W. Wang, W. Schnelle, S. Wirth, Y. Chen, S. T. B. Goennenwein and C. Felser,
	Nature Phys.  \textbf{14}, 1125 (2018).
	%Magnetic Weyl semimetals with broken time-reversal symmetry are expected to generate strong intrinsic anomalous Hall effects, due to their large Berry curvature. Here, we report a magnetic Weyl semimetal candidate, Co3Sn2S2, with a quasi-two-dimensional crystal structure consisting of stacked kagome lattices. Combining the kagome-lattice structure and the long-range out-of-plane ferromagnetic order of Co3Sn2S2, we expect that this material is an excellent candidate for observation of the quantum anomalous Hall state in the two-dimensional limit.
	
	\bibitem{Ye}
	L. Ye, M. Kang, J. Liu, F. von Cube, C. R. Wicker, T. Suzuki, C. Jozwiak, A. Bostwick, E. Rotenberg, D. C. Bell, L. Fu, R. Comin and J. G. Checkelsky,
	Nature  \textbf{555}, 638 (2018).
	%Theoretical work has predicted that kagome lattices may also host Dirac electronic states that could lead to topological and Chern insulating phases Here we study the d-electron kagome metal Fe3Sn2, which is designed to support bulk massive Dirac fermions in the presence of ferromagnetic order.
	%????????????? ?????? ???????????, ??? ??????? ?????? ????? ????? ????????? ??????????? ????????? ??????, ??????? ????? ???????? ? ?????????????? ? ??????????? ????? ?????. ????? ?? ??????? d-??????????? ??????-?????? Fe3Sn2, ??????? ???????????? ??? ????????? ????????? ????????? ????????? ?????? ? ??????????? ??????????????? ???????.
	
	\bibitem{Lin}
	%Flatbands and Emergent Ferromagnetic Ordering in Fe3Sn2 Kagome Lattices
	Zh. Lin, J.-H. Choi, Q. Zhang, W. Qin, S. Yi, P. Wang, L. Li, Y. Wang, H. Zhang, Zh. Sun, L. Wei, Sh. Zhang, T. Guo, Q. Lu, J.-H. Cho, Ch. Zeng, and Zh. Zhang,
	Phys. Rev. Lett. \textbf{121}, 096401 (2018).
	
	\bibitem{Boldrin}
	%Haydeeite: A spin-1/2 kagome ferromagnet The mineral haydeeite, alpha-MgCu3(OD)6Cl2, is a S=1/2 kagome ferromagnet that displays long-range magnetic order below TC=4.2 K with a strongly reduced moment
	D. Boldrin, B. Fak, M. Enderle, S. Bieri, J. Ollivier, S. Rols, P. Manuel, A. S. Wills,
	Phys. Rev. B \textbf{91}, 220408(R) (2015).
	
	\bibitem{Liu1}
	%Orbital-selective Dirac fermions and extremely flat bands in frustrated kagome-lattice metal CoSn
	Zh. Liu, M. Li, Q. Wang, G. Wang, Ch. Wen, K. Jiang, X. Lu, Sh. Yan, Y. Huang, D. Shen, J.-X. Yin, Z. Wang, Zh. Yin, H. Lei and Sh. Wang,
	Nature Comm.  \textbf{11}, 4002 (2020).
	
	
	%Co3V2O8 Joel S. Helton, Nicholas P. Butch, Daniel M. Pajerowski2, Sergei N. Barilo3 and Jeffrey W. Lynn Science Advances  01 May 2020 Vol. \textbf{6}, no. 18, eaay9709  neutron data reveal spin waves in the ferromagnetic ground state of the kagome staircase material Co3V2O8. While previous work has treated this material as quasi?two-dimensional, we find that an inherently three-dimensional description is needed to describe the spin wave spectrum throughout reciprocal space. At a higher temperature where Co3V2O8 displays an antiferromagnetic spin density wave structure, there are no well-defined spin wave excitations, with most of the spectral weight observed in broad diffuse scattering centered at the (0, 0.5, 0) antiferromagnetic Bragg peak.
	
	
	\bibitem{Guterding}
	%Prospect of quantum anomalous Hall and quantum spin Hall effect in doped kagome lattice Mott insulators
	D. Guterding, H. O. Jeschke and R. Valenti, Sci. Rep.  \textbf{6}, 25988 (2016)
	
	\bibitem{graph1}
	%Tunable Correlated Chern Insulator and Ferromagnetism in Trilayer Graphene/Boron Nitride Moir? Superlattice
	G. Chen, A. L. Sharpe, E. J. Fox, Y.-H. Zhang, S. Wang,
	L. Jiang, B. Lyu, H. Li, K. Watanabe, T. Taniguchi, Zh. Shi, T. Senthil, D. Goldhaber-Gordon, Y. Zhang, F. Wang, Nature  \textbf{579}, 56 (2020).
	%arXiv preprint arXiv:1905.06535 (2019).
	%Y.-H. Zhang, D. Mao, and T. Senthil Phys. Rev. Research 1, 033126 (2019).
	
	\bibitem{graph}
	%Intrinsic quantized anomalous Hall effect in a moire heterostructure
	M. Serlin, C. L. Tschirhart, H. Polshyn, Y. Zhang, J. Zhu, K. Watanabe, T. Taniguchi, L. Balents, A. F. Young, Science   \textbf{367},  900 (2020).
	
	\bibitem{Tsui}
	D.~C. Tsui, H.~L. Stormer, and A.~C. Gossard,
	Phys. Rev. Lett, \textbf{48}, 1559, (1982).
	
	\bibitem{Zhu}
	%Interaction-driven fractional quantum Hall state of hard-core bosons on kagome lattice at one-third filling
	W. Zhu, S. S. Gong, and D. N. Sheng,
	Phys. Rev. B \textbf{94}, 035129 (2016).
	
	
	\bibitem{Nielsen}
	%Local models of fractional quantum Hall states in lattices and physical implementation
	%https://www.nature.com/articles/ncomms3864
	A. E. B. Nielsen, G. Sierra and J. I. Cirac,
	Nature Comm. \textbf{4}, 2864 (2013).
	%The fractional quantum Hall effect is one of the most striking phenomena in condensed matter physics. It Here we propose a new way of constructing lattice Hamiltonians with local interactions and fractional quantum Hall like ground states. In particular, we obtain a spin 1/2 model with a bosonic Laughlin-like ground state, displaying a variety of topological features. We also demonstrate how such a model naturally emerges out of a Fermi?Hubbard-like model at half filling, in which the kinetic energy part possesses bands with non-zero Chern number ???????????? ???????????? ???? ???????? ? ????????????
	
	\bibitem{Gong}
	%Emergent Chiral Spin Liquid: Fractional Quantum Hall Effect in a Kagome Heisenberg Model
	Sh.-Sh. Gong, W. Zhu and D. N. Sheng,
	%The fractional quantum Hall effect (FQHE) realized in two-dimensional electron systems under a magnetic field is one of the most remarkable discoveries in condensed matter physics. Interestingly, it has been proposed that FQHE can also emerge in time-reversal invariant spin systems, known as the chiral spin liquid (CSL) characterized by the topological order and the emerging of the fractionalized quasiparticles. A CSL can naturally lead to the exotic superconductivity originating from the condense of anyonic quasiparticles. Although CSL was highly sought after for more than twenty years, it had never been found in a spin isotropic Heisenberg model or related materials. By developing a density-matrix renormalization group based method for adiabatically inserting flux, we discover a FQHE in a spin-12 isotropic kagome Heisenberg model. We identify this FQHE state as the long-sought CSL with a uniform chiral order spontaneously breaking time reversal symmetry, which is uniquely characterized by the half-integer quantized topological Chern number protected by a robust excitation gap. The CSL is found to be at the neighbor of the previously identified Z2 spin liquid, which may lead to an exotic quantum phase transition between two gapped topological spin liquids.
	Sci. Rep. \textbf{4}, 6317 (2014)
	
	
	\bibitem{Hasan} M. Z. Hasan, C. L. Kane, Rev. Mod. Phys. \textbf{82}, 3045 (2010).
	
	\bibitem{Haldane}
	F. D. M. Haldane, Phys. Rev. Lett. \textbf{61}, 2015 (1988).
	
	\bibitem{Neupert1}
	T. Neupert, L. Santos, C. Chamon, and C. Mudry,
	Phys. Rev. Lett. \textbf{106}, 236804 (2011).
	
	\bibitem{Zhang1}
	Y.-H. Zhang and T. Senthil, Phys. Rev. B \textbf{99}, 205150
	(2019).
	
	\bibitem{Zhang}
	Y.-H. Zhang and T. Senthil,
	Phys. Rev. B \textbf{102}, 115127 (2020).
	
	\bibitem{Hu}
	J. Hu, S.-Y. Xu, N. Ni, Zh. Mao,
	Ann. Review Mater. Research \textbf{49}, 207 (2019).
	
	
	\bibitem{Ando}
	Y. Ando, J. Phys. Soc. Jpn. \textbf{82}, 102001 (2013).
	
	\bibitem{Shvez}
	%?. ?. ???????, ?. ?. ????, ?. ?. ??????,  ?????? ? ???? \textbf{110}, 777 (2019).
	V. N. Men?shov, I. A. Shvets, E. V. Chulkov, JETP Letters \textbf{110}, 771 (2019).
	
	
	\bibitem{Wen2}
	X.-G. Wen, Adv. Phys. \textbf{44}, 405 (1995).
	
	\bibitem{Wen22}
	X.-G. Wen, {\it Quantum Field Theory of Many-body Systems}, Oxford University Press, Oxford, 2004.
	
	
	\bibitem{Moroz}
	S. Moroz, A. Prem, V. Gurarie, L. Radzihovsky, Phys. Rev. B \textbf{95}, 014508 (2017).
	
	
	\bibitem{Wen23}
	X.-G. Wen, Int. J. Mod. Phys. B\textbf{6}, 1711 (1992).
	
	\bibitem{Kitaev}
	A. Kitaev, Ann. Phys. (N. Y.) \textbf{321}, 2 (2006).
	
	
	
	\bibitem{Mao}
	Y.-H. Zhang, D. Mao, Phys. Rev. B \textbf{101}, 035122 (2020).
	
	\bibitem{Ma}
	R. Ma, Y.-Ch. He,  Phys. Rev. Research \textbf{2}, 033348 (2020).
	
	\bibitem{Parameswaran}
	S. A. Parameswaran, R. Roy, Sh.L. Sondhi, C. R. Physique \textbf{14}, 816 (2013).
	
	
	\bibitem{Scr4}
	V.Yu. Irkhin, Yu.N. Skryabin, Phys. Lett. A \textbf{383}, 2974  (2019).
	
	\bibitem{Hubbard}
	J. Hubbard,
	% Electron correlations in narrow energy bands.
	Proc. Roy. Soc. A \textbf{276}, 238 (1963). 
	
	\bibitem{Hubbard2}
	J. Hubbard, Proc. Roy. Soc. A \textbf{277}, 237 (1963).
	
	
	\bibitem{Hubbard3}
	J. Hubbard,
	% Electron correlations in narrow energy bands.
	Proc. Roy. Soc. A \textbf{281}, 401 (1964).
	
	\bibitem{Punk} M. Punk and S. Sachdev, Phys. Rev. B \textbf{85}, 195123 (2012).
	
	\bibitem{Kang}
	M. Kang, L. Ye, Sh. Fang, J.-Sh. You et al, 
	%A. Levitan, M. Han, J. I. Facio, C. Jozwiak, A. Bostwick, E. Rotenberg, M. K. Chan, R. D. McDonald, D. Graf, K. Kaznatcheev, E. Vescovo, D. C. Bell, E. Kaxiras, J. van den Brink, M. Richter, M. P. Ghimire, J. G. Checkelsky, R. Comin,
	%Dirac fermions and flat bands in the ideal kagome metal FeSn.
	Nat. Mater.\textbf{19}, 163 (2020).
	
	% X.G. Wen, Quantum Field Theory of Many-Body Systems --- From the Origin of   Sound to an Origin of Light and Electrons. Oxford University Press, 2004.
	
	
	\bibitem{Wen1}
	P.A. Lee, N. Nagaosa, and X.-G. Wen, Rev. Mod. Phys. \textbf{78}, 17  (2006).
	
	\bibitem{Scr2}
%	?. ?. ?????, ?. ?. ???????, 	??? \textbf{120}, 563 (2019).
V. Yu. Irkhin, Yu. N. Skryabin, Phys. Met. Metallogr. \textbf{120}, 513 (2019).
	
	
	
	\bibitem{Mei}
	E. Tang, J.-W. Mei, X.-G. Wen,
	Phys. Rev. Lett. \textbf{106}, 236802 (2011).
	
	
	\bibitem{Pollmann}
	%Quantum Spin Liquid with Even Ising Gauge Field Structure on Kagome Lattice
	Y.-Ch. Wang, X.-F. Zhang, F. Pollmann, M. Cheng, and Z. Y. Meng,
	Phys. Rev. Lett. \textbf{121}, 057202 (2018).
	
	\bibitem{Yanagi}
	Y. Yanagi, J. Ikeda, K. Fujiwara, K. Nomura, A. Tsukazaki, M.-T. Suzuki, Phys. Rev. B \textbf{103}, 205112 (2021).
	
	\bibitem{Zhou}
	H. Zhou, G. Chang, G. Wang, X. Gui, X. Xu, J.-X. Yin, Z. Guguchia, S. S. Zhang, T.-R. Chang, H. Lin, W. Xie, M. Z. Hasan, Sh. Jia,
	Phys.Rev. B\textbf{101}, 125121 (2020).
	
	\bibitem{Kassem}
	M. A. Kassem, Novel magnetic and electronic properties of kagome-lattice cobalt-shandites, PhD dissertation, Kyoto University. 2016.
	
	\bibitem{Yang}
	T. Y. Yang, Q. Wan, Y. H. Wang, M. Song, J. Tang, Z. W. Wang, H. Z. Lv, N. C. Plumb, M. Radovic, G. W. Wang, G. Y. Wang, Z. Sun, R. Yu, M. Shi, Y. M. Xiong, N. Xu,
	% Evidence of orbit-selective electronic kagome lattice with planar flat-band in correlated paramagnetic YCr6Ge6. Preprint at https://arxiv.org/abs/1906.07140
	arXiv:1906.07140 (2019).
	
	
	%\bibitem{633a} P. W.  Anderson, Science \textbf{235} 1196 (1987).
	%Int. J. Mod. Phys. B 25, 1 (2011).
	
	
	
	\bibitem{Fresar} R.~Fresard and P.~W\"{o}lfle,
	%Unified Slave Boson     Representation of Spin and Charge Degrees of Freedom for Strongly Correlated Fermi Systems,
	Int. J. Mod. Phys.~B~\textbf{6}, 685 (1992).
	
	\bibitem{I19} V. Yu. Irkhin, Phys. Lett. A \textbf{383}, 1506 (2019).
	
	
	\bibitem{654}
	V.Yu.~Irkhin, Yu.P.~Irkhin,
	%Many-electron operator approach in the solid state theory,
	phys. stat. sol. (b) 	\textbf{183}, 9 (1994).
	
	\bibitem{Ribeiro} T. C. Ribeiro, X.-G. Wen, Phys. Rev. B \textbf{74}, 155113 (2006).
	
	
	\bibitem{Scr}
%	?. ?. ?????, ?. ?. ???????, 	?????? ???? \textbf{106}, 161 (2017).
V. Yu. Irkhin, Yu. N. Skryabin, JETP Lett.  \textbf{106}, 167 (2017). 

	
	\bibitem{Song}
	X.-Y. Song, A. Vishwanath, Y.-H. Zhang, arXiv:2011.10044; Phys. Rev. B \textbf{103}, 165138 (2021).
	
	\bibitem{Vojta} M. Vojta, Rep. Prog. Phys. \textbf{81}, 064501 (2018).
	
	
	
	\bibitem{Scr3}
%	?. ?. ?????, ?. ?. ???????,	??? \textbf{121}, 115 (2020).
V. Yu. Irkhin, Yu. N. Skryabin, Phys. Met. Metallogr. \textbf{ 121}, 103 (2020). 
	
	
	\bibitem{Sachdev2}
	%Weak magnetism and non-Fermi liquids near heavy-fermion critical points
	T. Senthil, M. Vojta, and S. Sachdev,
	Phys. Rev. B \textbf{69}, 035111 (2004).
	
	
	\bibitem{Xu}
	G. Xu,  B. Lian,  and S.-C. Zhang,
	%Intrinsic Quantum Anomalous Hall Effect in the Kagome Lattice Cs2LiMn3F12.
	Phys. Rev. Lett. \textbf{115}, 186802 (2015).
	
	
	\bibitem{Bergholtz}
	%Topological Flat Band Models and Fractional Chern Insulators
	E. J. Bergholtz, Zh. Liu,
	%arXiv:1308.0343
	Int. J. Mod. Phys. B \textbf{27}, 1330017 (2013).
	
	\bibitem{Muechler}
	%Emerging chiral edge states from the confinement of a magnetic Weyl semimetal in Co3Sn2S2
	L. Muechler, E. Liu, J. Gayles, Q. Xu, C. Felser, Y. Sun, Phys. Rev. B \textbf{101}, 115106 (2020).

		
	\bibitem{Tanaka}
	%Topological Kagome Magnet Co3Sn2S2 Thin Flakes with High Electron Mobility and Large Anomalous Hall Effect
	M. Tanaka, Y. Fujishiro, M. Mogi, Y. Kaneko, T. Yokosawa, N. Kanazawa, S. Minami, T. Koretsune, R. Arita, S. Tarucha, M. Yamamoto, and Y. Tokura,
	Nano Lett.  \textbf{20},  7476 (2020).
		
	
	\bibitem{RMP} M. I. Katsnelson, V. Yu. Irkhin, L. Chioncel, A. I.
	Lichtenstein, and R. A. de Groot, Rev. Mod. Phys. \textbf{80}, 315 (2008).
	
	
	
	\bibitem{HgCr2Se4}
	G. Xu, H. Weng, Zh. Wang, X. Dai, Zh. Fang, Phys. Rev. Lett. \textbf{107}, 186806 (2011).
	
	
	\bibitem{Skr20}
	V.Yu. Irkhin, Yu. N. Skryabin,  JETP Lett. \textbf{111}, 230 (2020).
	
	
\end{thebibliography}
\end{document}